\documentclass[aps,prc,superscriptaddress,showpacs,nofootinbib,floatfix,twocolumn]{revtex4}

\usepackage{doi,graphicx,amssymb,amsmath,dcolumn}
\newcommand{\unit}[1]{\ensuremath{\, \mathrm{#1}}}
\newcommand{\etal}{\textit{et al.}}

\newcolumntype{d}[1]{D{.}{.}{#1} }
\begin{document}

\title{Simulating the Novel Phase Separation of a Rapid Proton Capture Ash Composition}
\author{M. E. Caplan}\email{mecaplan@indiana.edu}
\author{D. K. Berry}\email{dkberry@iu.edu}
\author{C. J. Horowitz}\email{horowit@indiana.edu}
\affiliation{Department of Physics and Nuclear Theory Center, Indiana University, Bloomington, IN 47405, USA}

\author{A. Cumming}\email{andrew.cumming@mcgill.ca}
\affiliation{Department of Physics and McGill Space Institute, McGill University,  Montreal QC, H3A 2T8 Canada} 
\author{R. Mckinven}\email{mckinven@astro.utoronto.ca}
\affiliation{Department of Astronomy \& Astrophysics, University of Toronto, ON M5S 3H4, Canada} 

\date{\today}
\begin{abstract}

Nucleosynthesis in the oceans of accreting neutron stars can produce novel mixtures of nuclides, whose composition is dependent on the exact astrophysical conditions. Many simulations have now been done to determine the nucleosynthesis products in the ocean, but the phase separation at the base of the ocean, which determines the composition of the crust, has not been as well studied. In this work, we simulate the phase separation of a composition, which was predicted to produce a crust enriched in light nuclei, in contrast with past work which predicts that crust is enriched in heavy nuclei. We perform molecular dynamics simulations of the phase separation of this mixture using the methods of Horowitz \etal\ (2007). We find good agreement with the predictions of Mckinven \etal\ (2016) for the phase separation of this mixture. Moreover, this supports their method as a computationally efficient alternative to molecular dynamics for calculating phase separation for a wider regime of astrophysical conditions.

\end{abstract}


\pacs{}

\maketitle

\section{Introduction}\label{s:phasesep_Intro}

Neutron stars in low mass X-ray binaries (LMXBs) accrete matter from a companion star, powering X-ray emissions. Accreted material arrives on the surface of the neutron star where it is heated and compressed under the weight of the later arriving material.
This can trigger episodes of unstable explosive nuclear burning, where the rapid proton capture process (rp-process) burns the hydrogen to a mixture of heavy isotopes, while under certain conditions, this material burns stably \cite{2003NuPhA.718..247S}.
 
The energy released from nuclear burning is approximately 5 MeV per nucleon for solar composition material, and will drive an order of magnitude increase in X-ray luminosity for 10-100 s. This transient is observed as an X-ray burst \cite{2008LRR....11...10C}. There is no one mixture of rp-process ash that forms in the accreted ocean. Hydrogen burns via the CNO cycle in the accreted ocean between bursts, while the rp-process operates during bursts, which creates a mixture which is uniquely determined by the astrophysical conditions, specifically, the accretion rate $\dot{m}$ and the helium mass fraction $Y$ in the accreted matter. 

As this material is further buried it will eventually freeze and form the solid outer crust of the star, a simple body centered cubic (bcc) lattice of nuclei embedded in a degenerate Fermi gas of electrons \cite{Horowitz:2009af}. Presently, the composition of the crust is not fully understood, though its composition and structure is important for calculating the thermal transport properties of the crust. 

To characterize the crust simply modelers often use $Q_{imp}$, the impurity parameter, which determines the electron-impurity scattering frequency, and is defined as the variance in charge of ions in the mixture

\begin{equation}
Q_{imp} \equiv  n_{ion}^{-1} \sum_{i} n_i ( Z_i - \langle Z \rangle)^2
\end{equation}

\noindent where $n_{ion}$ is the number density of ions (nuclei), and $n_i$ is the number density of the $i^{th}$ species which has charge $Z_i$, and $\langle Z \rangle$ is the average charge of all ions. Many crust models use $Q_{imp}$ as a proxy for detailed information about the crust composition \cite{crustcooling4, PhysRevLett.114.031102, Mckinven:2016zkg}.

Observations of LMXBs in quiescence find that the crust is well described with a low impurity parameter ($Q_{imp} < 10$). This is at odds with calculations of the nucleosynthetic yields from the rp-process, which predicts that the rp-ash will have a large impurity parameter ($Q_{imp} \sim 30-50 $) \cite{1999ApJ...524.1014S}. 
This discrepancy between observation and theory is currently an open problem, and motivates this work \cite{galloway2004, crustcooling4, PhysRevLett.114.031102, 2016arXiv160907155D, 0004-637X-833-2-186}.

Several solutions have been proposed, most notably, phase separation. Past work has shown that as the rp-ash mixture freezes, the lightest (lowest $Z$) nuclei may be preferentially retained in the liquid ocean, while heavier nuclei (high $Z$) are preferentially deposited into the solid crust \cite{Horowitz:2007hx}. This lowers the impurity parameter of the crust (relative to the ash) to be in agreement with observation. The ensuing enrichment of the ocean in light nuclei has also been the basis for suggestions that the ocean is convective, which allows it to maintain a uniform composition at the ocean-crust interface \cite{medin2011, Medin:2014zfa}. 


However, light nuclei cannot accumulate in the ocean indefinitely. It is possible that the accumulation of light nuclei shifts the equilibrium compositions of the solid and liquid, causing the formation of crystalline grains which are individually purified, such that the ocean alternates between depositing light nuclei and heavy nuclei, though this has yet to be explored. Alternatively, the light elements may remain in the ocean, and be burned and processed to heavier material in later bursts. Multizone models of nucleosynthesis in X-ray bursts may soon be able to explore these possibilities \cite{cyburt2016}, though again, this problem is presently unsolved.

Past work has studied the phase separation of two-component and three-component Coulomb plasmas, as well as the rp-ash, using molecular dynamics simulations \cite{WD2, PhysRevE.86.066413, phasesep}. In short, molecular dynamics is a computational technique for simulating N-body systems by iteratively solving Newton's equations of motion. 
Horowitz \etal\ (2007) began by preparing solid and liquid configurations, each with 13,824 ions with the same composition, which was the composition for the rp-ash predicted from nuclear burning simulations \cite{Horowitz:2007hx}. They then connected these two separate configurations on a face into one large configuration, containing 27,648 ions, and evolved that simulation at the melting temperature, so that the liquid region would not freeze and so the solid region would not melt. As the simulation evolved ions could diffuse freely between the solid region and the liquid region, changing their compositions. They found that the liquid region became enriched in light nuclei, while the solid became enriched in heavy nuclei. This is presently the conventional wisdom for the phase separation of the rp-ash. 

More recently, Medin \& Cumming developed a semi-analytic method for calculating equilibrium solid-liquid phase separation of classical multi-component plasmas \cite{Medin2010}. Their method, when applied to the composition studied by Horowitz \etal, was able to reproduce the results of the molecular dynamics, again finding the solid was enriched in heavy nuclei while the liquid was enriched in light nuclei. 

Most recently, Mckinven \etal\ applied the semi-analytic method of Medin \& Cumming to 32 different rp-ash compositions, produced for a range of astrophysical conditions for LMXBs \cite{Mckinven:2016zkg}. These rp-ash compositions were taken from nucleosynthesis calculations of Schatz \etal, Stevens \etal, and also included the composition of Horowitz \etal\ \cite{1999ApJ...524.1014S, 2014ApJ...791..106S, 2008PhRvL.101w1101G, 2003NuPhA.718..247S, phasesep}. Like Medin \& Cumming (2010), Mckinven \etal\ was able to reproduce the results of Horowitz \etal\, and found that most of these compositions phase separated in accordance with the conventional wisdom, where the light nuclei are retained in the ocean while the heaviest species settle into the crust. However, they found two composition for low accretion rates ($\dot{m}=0.1 \dot{m}_{Edd}$, $Y$=0.2752 and $\dot{m}=0.5 \dot{m}_{Edd}$, $Y$=0.55), with $\langle Z \rangle \approx 12$, where the equilibrium liquid was heavier than the solid (more specifically, the average charge of the liquid was greater than the average charge of the solid).

This sort of phase separation of rp-ash mixtures has not been seen before, and should be verified with molecular dynamics to confirm that the method of calculating phase separation from Medin \& Cumming (2010) is valid for compositions enriched in light isotopes, and also to confirm the validity of the result of Mckinven \etal.

Here, we seek to verify this result by simulating this rp-ash composition directly with molecular dynamics, following the methods of Horowitz \etal\ \cite{Horowitz:2007hx}. 
In short, we prepare a cubic simulation of crystal and a cubic simulation of liquid, both with identical initial compositions chosen to match the rp-ash, and combine them into one simulation so that nuclei may diffuse. After evolving this simulation for many timesteps, we find the abundances of each nuclear species in the solid and liquid region of the simulation and compare to the predictions of Mckinven \etal. In Sec. \ref{ss:MD} we review our molecular dynamics formalism. In Sec. \ref{ss:phasesep_init} we describe the initial conditions of our simulation and in Sec. \ref{ss:phasesep_simulation} we discuss the simulation itself. In Sec. \ref{ss:phasesep_results} we present results and we conclude in Sec. \ref{ss:phasesep_conc}.

\section{Molecular Dynamics}\label{ss:MD}

Our molecular dynamics formalism used here is the same as in our previous work, and is described in detail in Refs. \cite{Horowitz:2007hx, Horowitz:2009af}, and we review it briefly here. 

All simulations are performed using the Indiana University Molecular Dynamics (IUMD) CUDA-Fortran code, version 6.3.1, on the Big Red II supercomputer at Indiana University. IUMD v6.3.1 has previously been used for simulations of nuclear pasta in the inner crusts of neutron stars \cite{PhysRevC.91.065802, PhysRevC.91.065802, PhysRevC.90.055805, Horowitz:2015gda, PhysRevLett.114.031102, astromaterials}, though previous versions have been used to simulate ion mixtures \cite{2011arXiv1109.5095H, WD2, WD_PRL, soliddiffusion}. 

In our formalism, all nuclei are treated as point particles with charge $Z_i$. Two nuclei, separated by a distance $r_{ij}$ interact via a screened Coulomb potential

\begin{equation}
V(r_{ij})=\frac{Z_i Z_j e^2}{r_{ij}} \exp(-r_{ij}/\lambda).
\label{eq.V}
\end{equation}  

The exponential screening, due to the degenerate electron gas between ions, is calculated from the Thomas Fermi screening length

\begin{equation}
\lambda^{-1}=2\alpha^{1/2}k_F/\pi^{1/2}
\label{eq.lambda}
\end{equation}

\noindent where the electron Fermi momentum $k_F$ is $k_F=(3\pi^2n_e)^{1/3}$ and $\alpha$ is the fine structure constant. The electron density $n_e$ is equal to the ion charge density, $n_e=\langle Z\rangle n$, where $n$ is the ion density and $\langle Z\rangle$ is the average charge. The screening length is generally greater than the inter-ion spacing due to the high electron Fermi energy in the neutron star crust.

\section{Initial Conditions}\label{ss:phasesep_init}

\begin{figure*}[t!]
\centering
\includegraphics[width=0.95\textwidth]{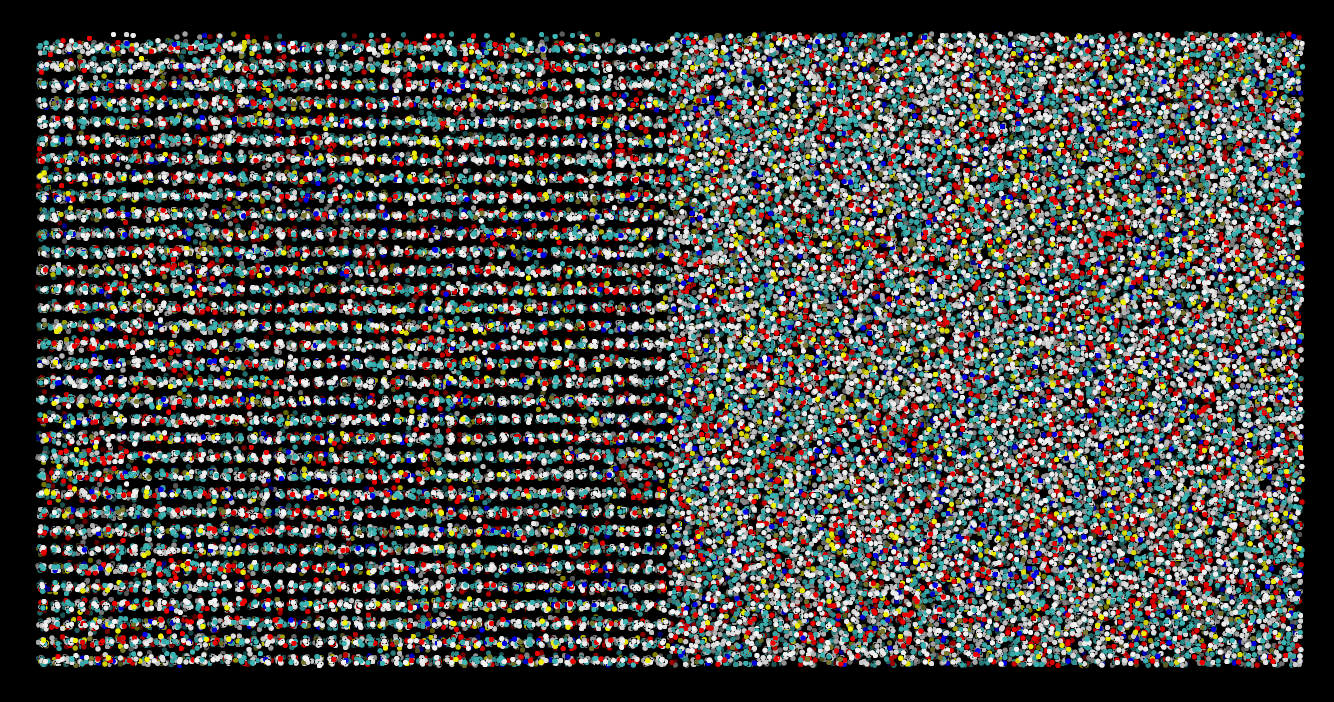}
\caption{\label{fig:phasesep1} (Color online) Initial conditions of our simulation. Each point represents one nucleus. Additionally, the simulation is 3D, but an orthographic projection is used here for clarity; on the left, the solid lattice is easily resolved, while the right is a liquid. The apparent clustering on lattice sites in the solid is due to the distribution of nuclei in the third dimension; the simulation is approximately a bcc lattice and each point is near a lattice site. The colors are dimmed in an effort to illustrate their depth in the field, and are also used to identify ion species.
 In order of abundance: $Z=12$ white, $Z=10$ cyan, $Z=8$ red, $Z=18$ blue, $Z=11$ in tan, remaining six species (totaling 5\% of all ions) shown in yellow. Compare to our final configuration, Fig \ref{fig:phasesep2}. } 
\end{figure*}

In this section we discuss the preparation of the initial configuration for our simulation. We follow the procedure of Horowitz \etal\ (2007) to create a configuration of 55,296 nuclei which is half solid and half liquid, which is shown in Fig. \ref{fig:phasesep1}.

To begin, we use the composition shown in Tab. \ref{tab:comp1}, taken from Schatz \etal \cite{1999ApJ...524.1014S}, and exclude carbon and all lighter nuclei. This is to be consistent with Mckinven \etal 

\begin{table}[h!]
\centering
\caption{Composition of the rp-ash for accretion rate $\dot{m}=0.1 \dot{m}_{Edd}$ with helium fraction $Y$=0.2752. The table lists the fractional abundance (by number) $n$ for each nuclide the charge of that species $Z$, with mass $A$.}\label{tab:comp1}

\begin{tabular}{lll}
\toprule
n     \hspace{3em} & Z  & A  \\ \hline
0.457  & 12 & 24 \\
0.279  & 10 & 20 \\
0.133  & 8  & 16 \\
0.038  & 18 & 40 \\
0.037  & 11 & 23 \\
0.023  & 22 & 52 \\
0.020  & 9  & 19 \\
0.005  & 24 & 56 \\
0.004  & 20 & 48 \\
0.002  & 21 & 49 \\
0.002  & 23 & 53 \\ 
\botrule
\end{tabular}
\end{table}

\noindent In reality, the burning produces many more different species than shown here. For example, there are often multiple isotopes of each element, and many species with abundances less than $10^{-3}$. For simplicity, we take all isotopes of the same element to have the same mass, chosen to be the most abundant isotope of that species. Additionally, we exclude all species less abundant than $10^{-3}$.

We use this to initialize a simulation of 3,456 nuclei. This number is chosen for two reasons. First, it is convenient for equilibrating a crystal. A bcc lattice has 2 nuclei per unit cell, so 3,456 nuclei will fit exactly into a crystal lattice of $12 \times 12 \times 12$ unit cells. Second, it is a small number of nuclei, and so simulations will run quickly. 

Simulations were run using the parameters in Tab. \ref{tab:run_m0010_small} to find the melting/freezing temperature of this mixture. The choice of a density of $7.18 \times 10^{-5} \unit{ fm}^{-3}$ deserves comment. This number density is significantly higher than what would be expected at the ocean-crust interface. It was chosen to be consistent with previous molecular dynamics simulations. Nevertheless, this will not effect the phase separation. In general, the literature reports the melting temperature of a mixture using the dimensionless Coulomb coupling parameter $\Gamma_{crit}$. For a one component mixture, 

\begin{equation}
\Gamma = \frac{e^2 Z^2}{aT} 
\end{equation}

\noindent where $a=(3/4 \pi n)^{1/3}$ is the Wigner-Seitz radius. Even though we are not simulating at a physically realistic density, our simulations are run at the physical melting temperature (i.e. at $\Gamma_{crit}$), and the results of molecular dynamics simulations can be arbitrarily rescaled to any specific physical density or temperature.

\begin{table}[h!]
\centering
\caption{Parameters of our simulation for the composition in Tab. \ref{tab:comp1}, including ion number density $n$, initial temperature $T_0$, Coulomb screening length $\lambda$, and timestep $dt$.  }\label{tab:run_m0010_small}
\begin{tabular}{lr}
\toprule
Parameter & Value \\ \hline
$T_0$ & 0.040 MeV \\
$n$  \hspace{3em}   & $7.18 \times 10^{-5}$ fm$^{-3}$   \\ 
$\lambda$  &  35.8516 fm \\ 
$dt$  & 25 fm/c  \\ 
\botrule
\end{tabular}
\end{table}

Simulations were started at high temperatures and lowered until the mixture froze. Approximate values of the melting temperature can be calculated using $\Gamma_s = 273.5$ ($T = 0.0457$ MeV) from Mckinven \etal. 
 However, to make an MD simulation freeze in a feasible length of time one must supercool the simulation, and thus freezing occurs at temperatures below (perhaps 10-20\%) the true freezing temperature. 

This simulation of 3,456 nuclei was evolved for $2,240 \times 10^{6}$ fm/c ($89.6 \times 10^{6}$ timesteps), and trial and error was used to find the freezing temperature. To briefly describe the thermodynamic history, it was initialized from random positions at a temperature of 0.040 MeV, and heated up to 0.130 MeV to equilibrate. It was cooled to 0.040 MeV where it froze, and then reheated to 0.060 MeV where it melted. The configuration was then cooled back to 0.036 MeV and froze again. 

This crystal at 0.036 MeV was then `tiled' to produce a larger configuration with 27,648 nuclei. This was prepared by translating the crystal formed to make a $2 \times 2 \times 2$ volume, thus containing our original simulation 8 times. This system was then equilibrated for an additional $400 \times 10^{6}$ fm/c ($16 \times 10^{6}$ timesteps) and heated from 0.036 MeV up to 0.042 MeV, which was believed to be nearer, and perhaps slightly above, the true melting point.

Additionally, using a liquid configuration of 3,456 nuclei at $T = 0.0413$ MeV generated in our earlier simulation, another configuration was assembled in the same manner, so that we could have a liquid with 27,648 nuclei with an identical composition to the crystal. This configuration was then equilibrated for $50 \times 10^{6}$ fm/c ($2 \times 10^{6}$ timesteps) at constant temperature, 0.042 MeV. 

These two configurations were then joined on one face to prepare a final configuration with 55,296 nuclei, which was shown in Fig. \ref{fig:phasesep1}.

\section{Simulation}\label{ss:phasesep_simulation}

\begin{figure*}[t!]
\centering
\includegraphics[width=0.85\textwidth]{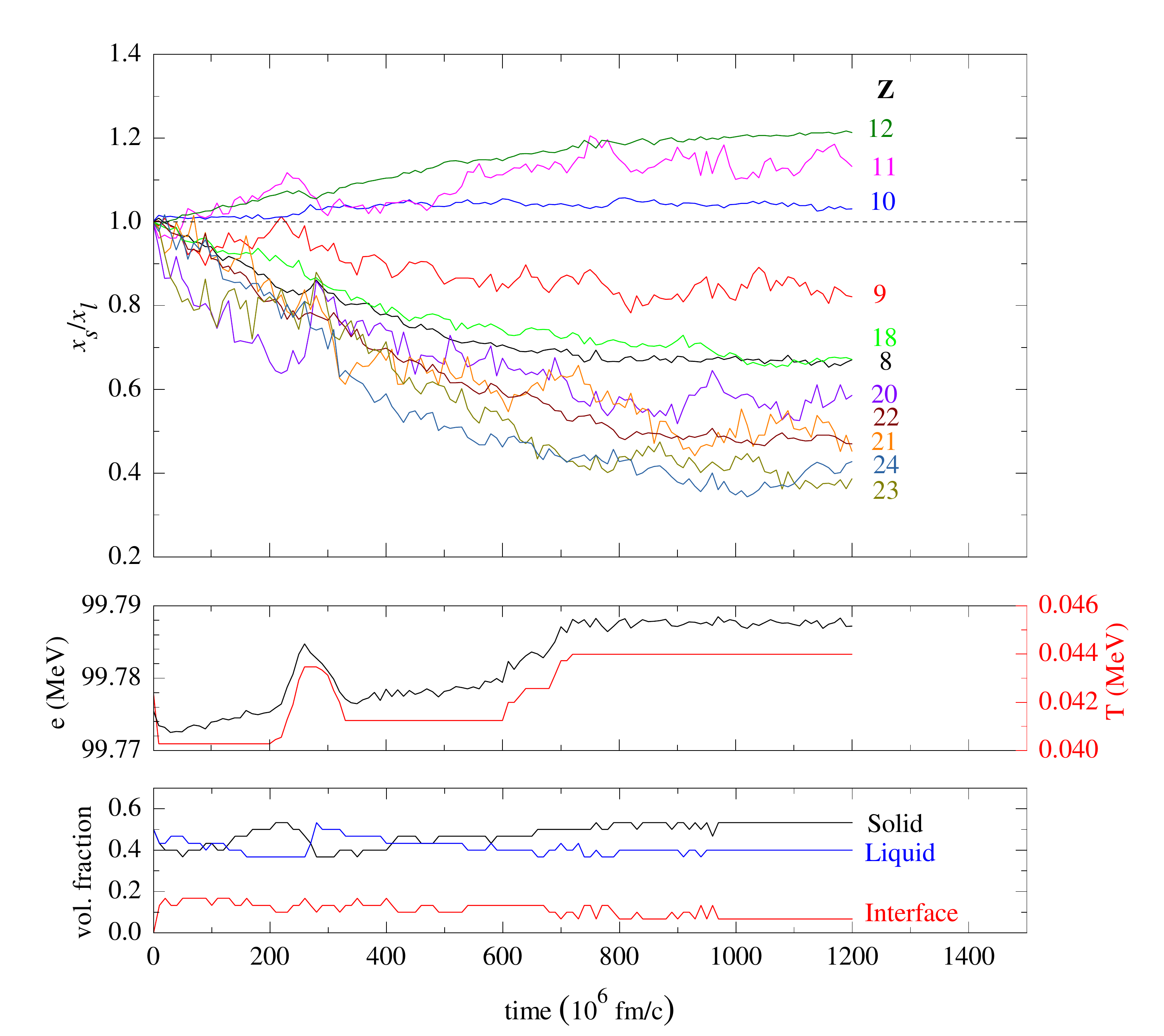}
\caption{\label{fig:m0010_evolution} (Color online) Evolution of our simulation. (Top) Abundance ratios in the solid and liquid region are shown as a function of time. (Middle) The temperature history for our simulation. The set point temperature (red) was adjusted manually in order to keep the simulation half liquid and half solid by volume. We also show the average total energy per nucleus (black). (Bottom) The volume fraction of the solid region, liquid region, and approximate thickness of the interface between them. These were determined by inspection, as described in the text.}
\end{figure*}

The half-solid/half-liquid configuration of 55,296 nuclei was evolved using the same parameters shown in Tab. \ref{tab:run_m0010_small} for $1,200 \times 10^6$ fm/c ($48\times 10^6$ timesteps). The initial temperature was 0.041 MeV. The temperature was chosen to be as close to the melting/freezing temperature as possible, and was adjusted periodically in order to ensure that half of the simulation volume remained solid while half remained liquid. To verify that the simulation had equal parts solid and liquid, we partitioned the simulation volume into 30 equal sized `slices' along the long axis every $10^7$ fm/c, and determined which slices were solid, liquid, or interfacial by inspection. A full thermodynamic history is shown in the middle panel of Fig \ref{fig:m0010_evolution}.




We present abundances for each nuclear species as the ratio $x_s / x_l$ where $x_s$ is the number fraction of nuclei of that species found in the solid and $x_l$ is the number fraction of nuclei of that species found in the liquid. Thus, $x_s / x_l = 1$ implies equal abundance in the solid and liquid, $x_s / x_l > 1$ implies the species is enriched in the solid and depleted in the liquid, whereas $x_s / x_l < 1$ implies the species is depleted in the solid and enriched in the liquid. We track the evolution of the abundances in the top panel of Fig. \ref{fig:m0010_evolution}. 

Initially, both the solid and the liquid have identical compositions, so $x_s / x_l = 1 $ for all species. Within $100\times10^6$ fm/c ($4\times10^6$ MD timesteps), the compositions have begun to differentiate, enhancing the solid in nuclei with $Z=10-12$ while depleting it in all other species. This trend continues out until about $900 \times 10^6$ fm/c ($36 \times 10^6$ MD timesteps), at which point the nuclear abundances stop changing. The simulation was run for a total of $1,200\times10^6$ fm/c ($48\times10^6$ MD timesteps), and the abundances in the solid and liquid did not change appreciably between $900\times10^6$ fm/c and  $1,200\times10^6$ fm/c, so we argue that this system has equilibrated.

\section{Results}\label{ss:phasesep_results}

We now study the final configuration and compare our abundances to Mckinven \etal\




From Fig. \ref{fig:m0010_evolution}, we find good qualitative agreement with Mckinven \etal\ The solid in our simulation is enhanced in $Z=10-12$ nuclei but is depleted in all other species, consistent with Mckinven \etal\ Additionally, relative abundances show the same pattern. For example, our solid is most strongly enhanced in $Z=12$ nuclei, less strongly enhanced in $Z=11$ nuclei, and least strongly enhanced in $Z=10$ nuclei, consistent with Mckinven \etal\ The same pattern also holds for the depleted species. 

However, our phase separation is not as strong. Mckinven \etal\ predicts that the enhancement of $Z=10-12$ nuclei in the solid should be approximately $1.5\times$ greater, while the depletion in nuclei $Z>20$ should be an order of magnitude stronger. To understand this disagreement, we must examine the structure of the final configuration of our simulation.

\begin{figure*}
\centering
\includegraphics[width=0.95\textwidth]{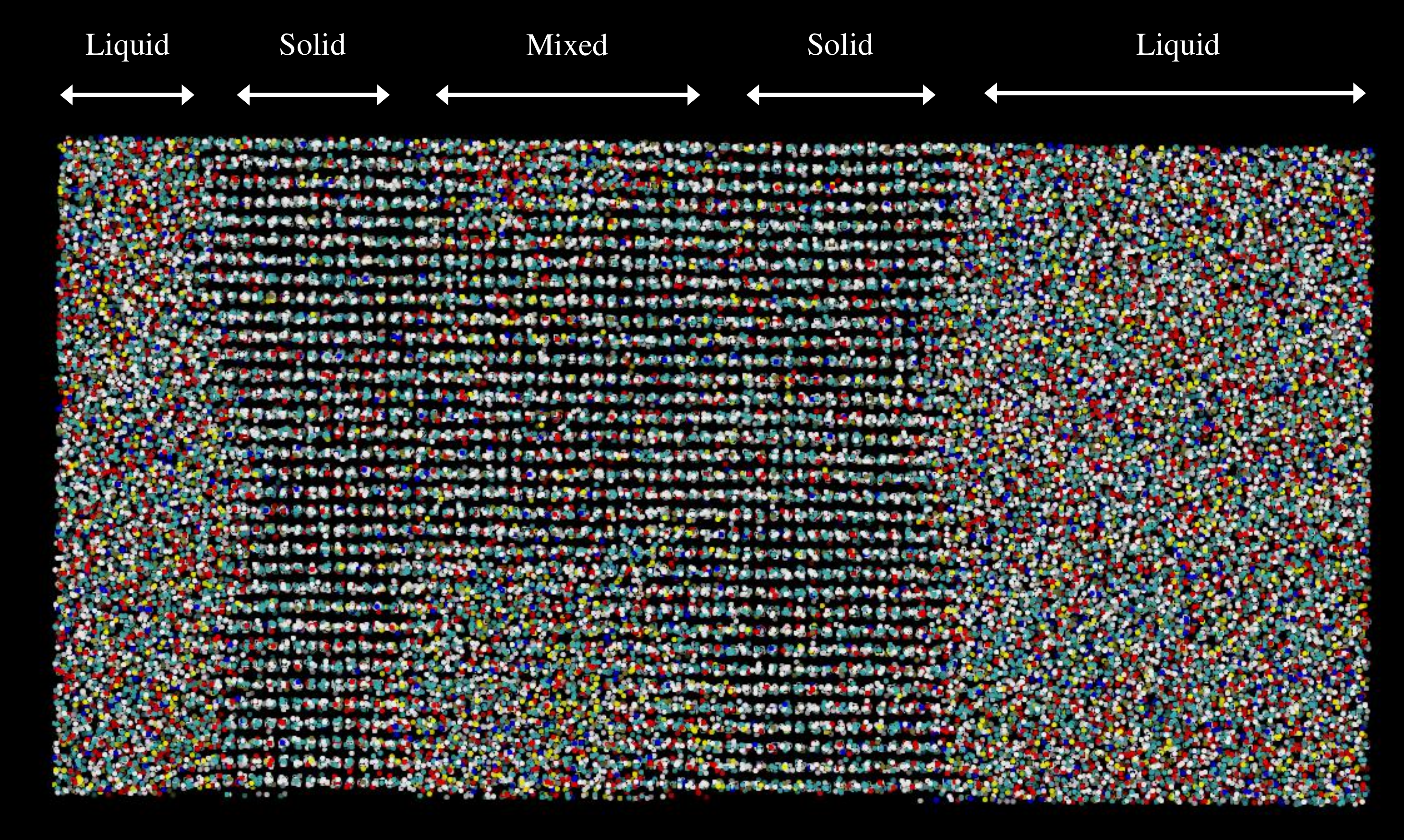}
\caption{\label{fig:phasesep2} (Color online) Final configuration of our simulation, after $1,200 \times 10^6$ fm/c; the same projection and color scheme as Fig \ref{fig:phasesep1} is used. Notice that the solid region of the simulation has shifted slightly, and that an amorphous region has formed inside of it. This `mixed' region is discussed in the text.}
\includegraphics[width=0.80\textwidth]{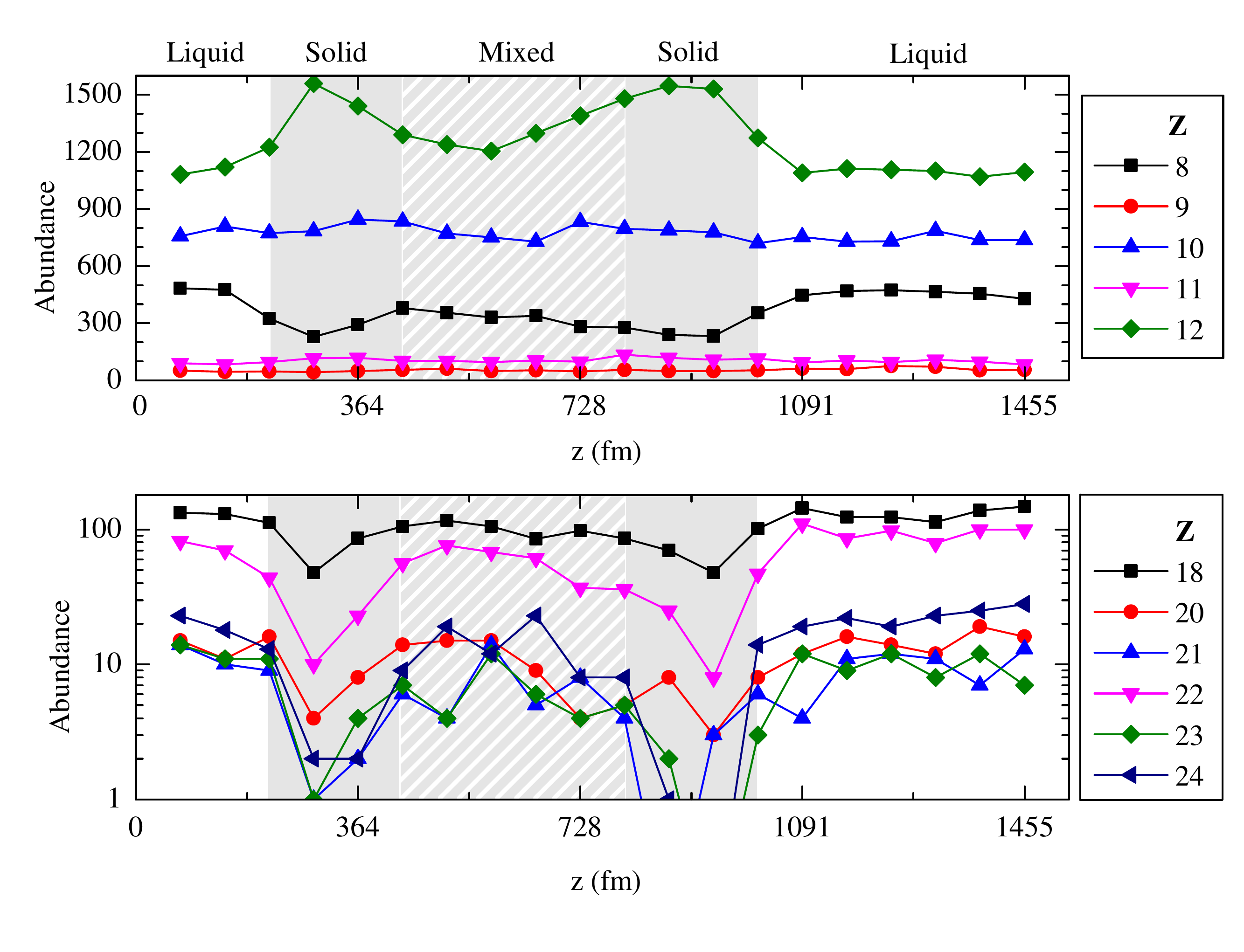}
\caption{\label{fig:slices} Total number of nuclei for each of the 20 subdomains along the z-axis of our simulation. The z-axis shown here is the same as shown in Fig. \ref{fig:phasesep2}.}
\end{figure*}

Several features emerge when the configuration is plotted, shown in Fig. \ref{fig:phasesep2}.
To begin, boundaries between the solid and liquid have shifted slightly, forming a small liquid region near the left face of the simulation. This is due to the periodic boundary conditions, so this liquid region is the same as the one on the right face. This shifting of the boundaries of the solid and liquid was accounted for in our calculations of the abundances, and does not effect $x_s / x_l$ as presented in Fig. \ref{fig:m0010_evolution}.

However, it is apparent that the solid region is not a uniform lattice. An amorphous region has formed inside the lattice. We conjecture that this region is the result of phase separation as well, where those nuclei too far from the faces of the solid region to diffuse outward instead diffused inward, creating a pocket of liquid within the solid.

We study the abundances along the length of the simulation by counting the number of nuclei in `slices.' The simulation dimensions are $727.5 \times 727.5 \times 1455$ fm. We partition along the long axis (z-axis) into 20 subdomains of equal thickness, such that each region contains approximately 2750 nuclei (this is similar to our method of finding the abundances in Fig. \ref{fig:m0010_evolution}). Then, we count the number of nuclei of each species in each subdomain. This is shown in Fig. \ref{fig:slices}.

When plotted this way, it becomes clear that a significant subvolume of what we call the solid region has a distinct composition. Abundances of several species in the center of the solid, particularly $Z=12$, are between what is seen in the regions that are more obviously solid and liquid. Because this subvolume is amorphous and has abundances between what would be expected for a solid or liquid, we take to calling this region the `mixed' region. 

Viewed this way, we notice that for most species where $Z>20$, there is an order of magnitude difference in abundance between the clearly solid slices, and the mixed or liquid slices. Knowing this, we can reanalyze our abundance data and exclude the mixed region. 

\begin{figure}
\centering
\includegraphics[width=0.48\textwidth]{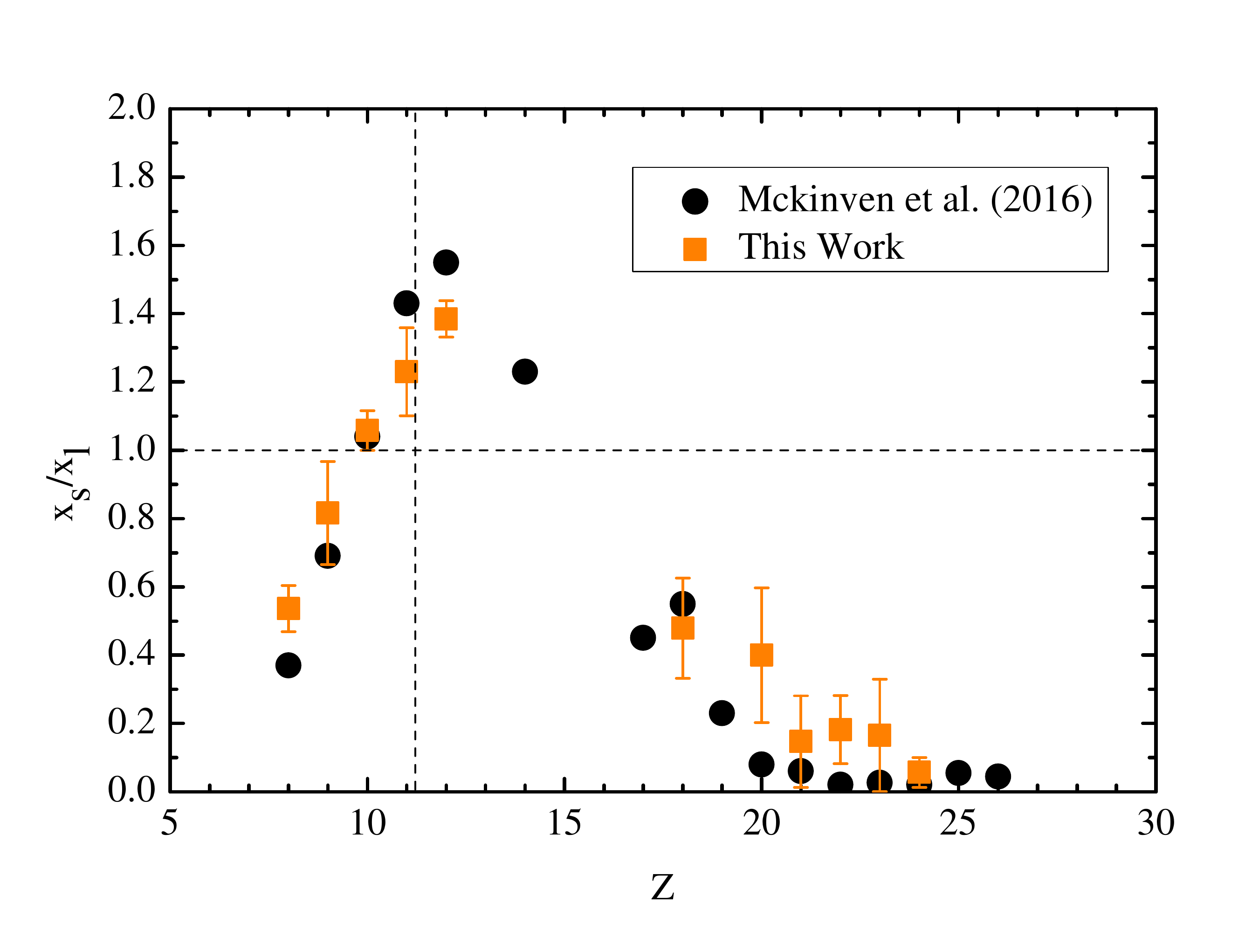}
\caption{\label{fig:xsxl_slices} Abundances in our simulation, excluding interfaces and the mixed subvolume. Note that the data from Mckinven \etal\ includes more species than this work, as we exclude all elements which are abundant to less than $10^{-3}$. }
\end{figure}

We recalculate $x_s / x_l$ for the solid and liquid, but this time we exclude data from the interfacial slices, and the mixed slices. Specifically, only four solid subdomains were included; these were the two solid subdomains on each side of the mixed region.

We show $x_s / x_l$ for each species in our simulation in Fig. \ref{fig:xsxl_slices}. Uncertainties are obtained from the standard errors of abundances for each of the four subdomain included in the analysis. With this, we find much better agreement between our molecular dynamics simulation and Mckinven \etal\ Notably, we find further enhancement in the solid of the $Z=$11 and 12 nuclei, while the depletion in species $Z>18$ is much stronger, and for many species we agree within error. 

\section{Discussion}\label{ss:phasesep_conc}

Fig. \ref{fig:xsxl_slices} is compelling, and we find the molecular dynamics to be in agreement with Mckinven \etal, and so we conclude that their method is accurate for the purposes of calculating phase separation of these rp-ash mixtures. 

Past work (in particular, Horowitz \etal\ 2007) studied compositions that were enriched in heavy nuclei, and found that the solid favors high $Z$ (i.e. $x_s / x_l > 1$ for $Z > \langle Z \rangle$). However, that work concluded that the high $Z$ nuclei form the lattice while the low $Z$ nuclei remain in the ocean, which was taken to be a general feature of rp-ash phase separation. 

With this result, we conclude (in agreement with Mckinven \etal) that there is a second general type of phase separation, which applies to rp-ash mixtures with a low average charge ($\langle Z \rangle \lesssim 13$) and an abundance of light nuclei. In these mixtures, the most abundant nuclei in the mixture form the solid, including other nuclei most similar in charge, while all other nuclei are enhanced in the liquid. 


This result also has a simple physical interpretation. The most abundant species sets the interparticle spacing in the lattice, so those species similar in charge can fit on lattice sites without disrupting the spacing and long range order, while those that are substantially different in charge will not fit, and be expelled from the solid. 

This may have implications for the time evolution of the ocean and crust. Given that we have now identified compositions which form a light crust following phase separation, it may be possible to form a multicomponent solid in the crust. As a high $Z$ crust forms under `normal conditions,' the ocean may accumulate light nuclei. This enrichment may eventually cause the equilibrium composition of the solid to shift so that the ocean then deposits a crystal of light nuclei. This could possibly be a mechanism for producing a polycrystalline crust with a low $Q_{imp}$, though the time dependent phase separation of these mixtures has not yet been explored. 

Now that molecular dynamics has shown the qualitative existence of this type of phase separation, future molecular dynamics simulations could seek to resolve the boundary between the two kinds of phase separation. For example, Mckinven \etal\ reports two compositions which have quantitatively similar abundances ($\dot{m}=0.5$, $Y=0.55$ and $\dot{m}=1.0$, $Y=0.55$), where one composition phase separates to produce a heavy solid while the other composition phase separates to create a light solid. Molecular dynamics simulations to verify these predictions could be useful for understanding the transition between these two types of phase separation.

The presence of the mixed phase in our simulation was unique, as no such thing has been reported in previous molecular dynamics simulations of phase separation. Perhaps this is due to the small simulation sizes previously used. For example, in this work we used twice as many particles as Horowitz \etal\ While not erroneous, it was a nuisance, and future work may seek to avoid producing the mixed phase structures. 

To prevent the formation of a `mixed' region, or liquid subvolume, in the future with larger molecular dynamics simulations, we make the following suggestions. First, it may be useful to try thinner solid and liquid regions, so that nuclei can have a larger amount of surface area per volume to diffuse out from. For example, Horowitz \etal\ used a cubic simulation volume, so that their solid and liquid regions were each $727 \times 727 \times 364$ fm on a side, while ours were cubic, $727 \times 727 \times 727$ fm on a side \cite{Horowitz:2008jt}. Alternatively, it may be sensible to initialize the simulation using the phase separated compositions of the solid and liquid predicted from the models, and then observe that the composition remains stable and does not phase separate further. 

The agreement of this work with Mckinven \etal\ does not mean that the composition of the crust is now known. The composition of the rp-ash which forms from burning is still not fully understood, the ocean may still be convective, and there may be time dependence to the phase separation. Future work should seek to generalize the approach of Mckinven \etal\ to a time dependent ocean composition in order to study the formation of the crust, and future molecular dynamics simulations may seek to study the phase separation with a convective ocean, though this may require significantly larger simulations than those discussed here. To truly determine the composition of the crust of a neutron star in an accreting binary, one may need to couple the methods of Mckinven \etal\ with multizone nuclear reaction networks.


\begin{acknowledgments}

This research was supported in part by Lilly Endowment, Inc., through its support for the Indiana University Pervasive Technology Institute, and in part by the Indiana METACyt Initiative. The Indiana METACyt Initiative at IU was also supported in part by Lilly Endowment, Inc.

This research was supported in part by DOE grants DE-FG02-87ER40365 (Indiana University) and DE-SC0008808 (NUCLEI SciDAC Collaboration). AC is supported by an NSERC Discovery Grant.

\end{acknowledgments}

\bibliography{references}

\end{document}